\documentclass{rsproca}
\usepackage{natbib}

\begin{document}

\title{The Role of Turbulence in Coronal Heating and Solar Wind Expansion}

\author{Steven R.\  Cranmer$^{1,2}$,
Mahboubeh Asgari-Targhi$^{1}$,
Mari Paz Miralles$^{1}$,
John C.\  Raymond$^{1}$,
Leonard Strachan$^{1,3}$,
Hui Tian$^{1}$,
and
Lauren N.\  Woolsey$^{1}$}

\address{$^{1}$Harvard-Smithsonian Center for Astrophysics,
Cambridge, MA 02138, USA\\
$^{2}$Laboratory for Atmospheric and Space Physics,
Department of Astrophysical and Planetary Sciences,
University of Colorado, Boulder, CO 8030, USA (as of January 2015)\\
$^{3}$Naval Research Laboratory, 
Space Science Division, Code 7684,
Washington, DC 20375, USA (current affiliation)}

\subject{astrophysics, solar physics}

\keywords{heliosphere, plasma physics, solar atmosphere,
solar corona, solar wind, turbulence}

\corres{Steven R.\  Cranmer\\
\email{scranmer@cfa.harvard.edu}}

\begin{abstract}
Plasma in the Sun's hot corona expands into the heliosphere as a
supersonic and highly magnetized solar wind.
This paper provides an overview of our current understanding
of how the corona is heated and how the solar wind is accelerated.
Recent models of magnetohydrodynamic turbulence have progressed
to the point of successfully predicting many observed properties
of this complex, multi-scale system.
However, it is not clear whether the heating in open-field regions
comes mainly from the dissipation of turbulent fluctuations that
are launched from the solar surface, or whether the chaotic
``magnetic carpet'' in the low corona energizes the system via
magnetic reconnection.
To help pin down the physics, we also review some key observational
results from ultraviolet spectroscopy of the collisionless outer
corona.
\end{abstract}


\begin{fmtext}
%
\end{fmtext}
\maketitle

\section{Introduction}
\label{sec:intro}

The origin of the Sun's supersonic and turbulent solar wind is
linked intimately to the existence of the high-temperature
($T > 10^{6}$~K) solar corona.
There is still no comprehensive understanding of the physical
processes that generate both the coronal heating and
the solar wind's acceleration.
The early history of these unsolved problems reaches back more than
a century \citep[see][]{Hf91,P01}.
\citet{P58} combined many of the existing observational clues to
synthesize a single-fluid model of time-steady acceleration, in which
gravity was counteracted by the large gas pressure gradient of the
million-degree corona.
Just a few years later, {\em Mariner~2} confirmed the existence of
a continuous solar wind with roughly the properties predicted by
Parker \citep[e.g.,][]{NS66}.

Early {\em in~situ} measurements showed a range of plasma
conditions at 1~AU, including:
(1) a slow, dense, and highly variable component
($v \approx 300$ km~s$^{-1}$) that dominates in the ecliptic plane,
(2) calmer, but wave-filled high-speed streams ($v \approx 700$ km~s$^{-1}$),
and (3) occasional explosive disruptions (``coronal mass ejections'')
accompanied by distorted fields and geomagnetic storms.
Each of these phenomena was later found to correlate reasonably well
with distinct solar features identified via remote sensing---e.g.,
(1) streamers and ``quiet Sun,''
(2) coronal holes, and
(3) flares and active regions, respectively---but there are
frequent exceptions to these general associations.
The different coronal source regions have dramatically distinct
appearances, plasma properties, and spatial/temporal variability.
Thus, it is likely that the heliospheric plasma that comes from these
regions is energized by {\em different} combinations of mechanisms.
In other words, there is probably not one single piece of physics that
will ``solve the coronal heating problem'' once and for all.

Much of the remainder of this paper will review the current state
of debate regarding which physical processes are most active in the
acceleration regions of the slow and fast solar wind.
However, we do not want to overstate the degree of controversy.
It is important to note that nearly everyone agrees on the
following basic elements:
\begin{enumerate}
\item
There is more than enough energy in the Sun's convective overturning
motions to power the corona and solar wind.
\item
Convective granulation acts as a major ``lower boundary condition''
for the injection of energy into the topologically complex magnetic field.
\item
Magnetic energy is (somehow) transported up to coronal heights
and becomes organized into small-scale, nonlinear, and largely
field-aligned plasma structures.
\item
Magnetic and/or kinetic energy is (somehow) converted irreversibly to
thermal energy by a combination of Coulomb collisions and
collisionless wave-particle interactions.
This conversion probably occurs most efficiently in regions with
sharp spatial gradients of the plasma properties.
\item
The high coronal temperature gives rise to a strong divergence
of the pressure tensor, which in turn drives the dominant outward
acceleration of the solar wind.
Some other acceleration components are probably necessary to
accelerate the fastest solar wind streams from coronal holes.
\end{enumerate}
This paper presents a review of the authors' perspective on the
above ideas and is not a comprehensive overview of the literature.

In Section \ref{sec:photo} we discuss the importance of understanding
the photospheric lower boundary condition in various turbulent
coronal heating scenarios.
Section \ref{sec:wtd} summarizes models based on the dissipation
of waves and turbulent eddies along open field lines.
Section \ref{sec:rlo} summarizes models based on magnetic
reconnection and the opening up of closed loops.
Section \ref{sec:uvcs} reviews observations that point to efficient
collisionless kinetic processes acting in the solar wind's
acceleration region.
Section \ref{sec:conc} concludes with a brief summary of how
studying turbulence in the solar corona has helped improve our
broader understanding of other issues in astrophysics and
plasma physics.

\section{Photospheric energy sources}
\label{sec:photo}

The Sun's visible surface is threaded ubiquitously by magnetic fields,
and these fields are believed to expand out through the solar corona
and fill the heliosphere.
However, accurate and regular measurements of the magnetic field are
possible only at the photosphere (with spectroscopic Zeeman splitting
and other polarization diagnostics) and in distant regions
sampled by {\em in~situ} magnetometer probes.
Our knowledge of the coronal magnetic field comes mainly from
model-based extrapolation techniques \citep[e.g.,][]{HS86,Wi05}.

On the largest spatial scales (of order 0.1 to 1 solar radii, $R_{\odot}$)
the photospheric field is organized into low-order multipole components
and active regions that are driven by the solar dynamo.
On smaller scales of roughly $10^{-2} \, R_{\odot}$, the magnetic field
becomes fragmented into the so-called supergranular network
\citep{Le62,Ga76}.
Below that is the convective-scale granulation, with rapidly evolving
cell-like structures ranging between $10^{-4}$ and
$10^{-3} \, R_{\odot}$ in size.
It is still not known whether the supergranulation is a manifestation
of convective instability (i.e., just a ``deeper'' kind of granulation),
or whether it arises from other processes \citep[see, e.g.,][]{Ra03,HS14}.

\begin{figure}[!t]
\centering\includegraphics[width=4.0in]{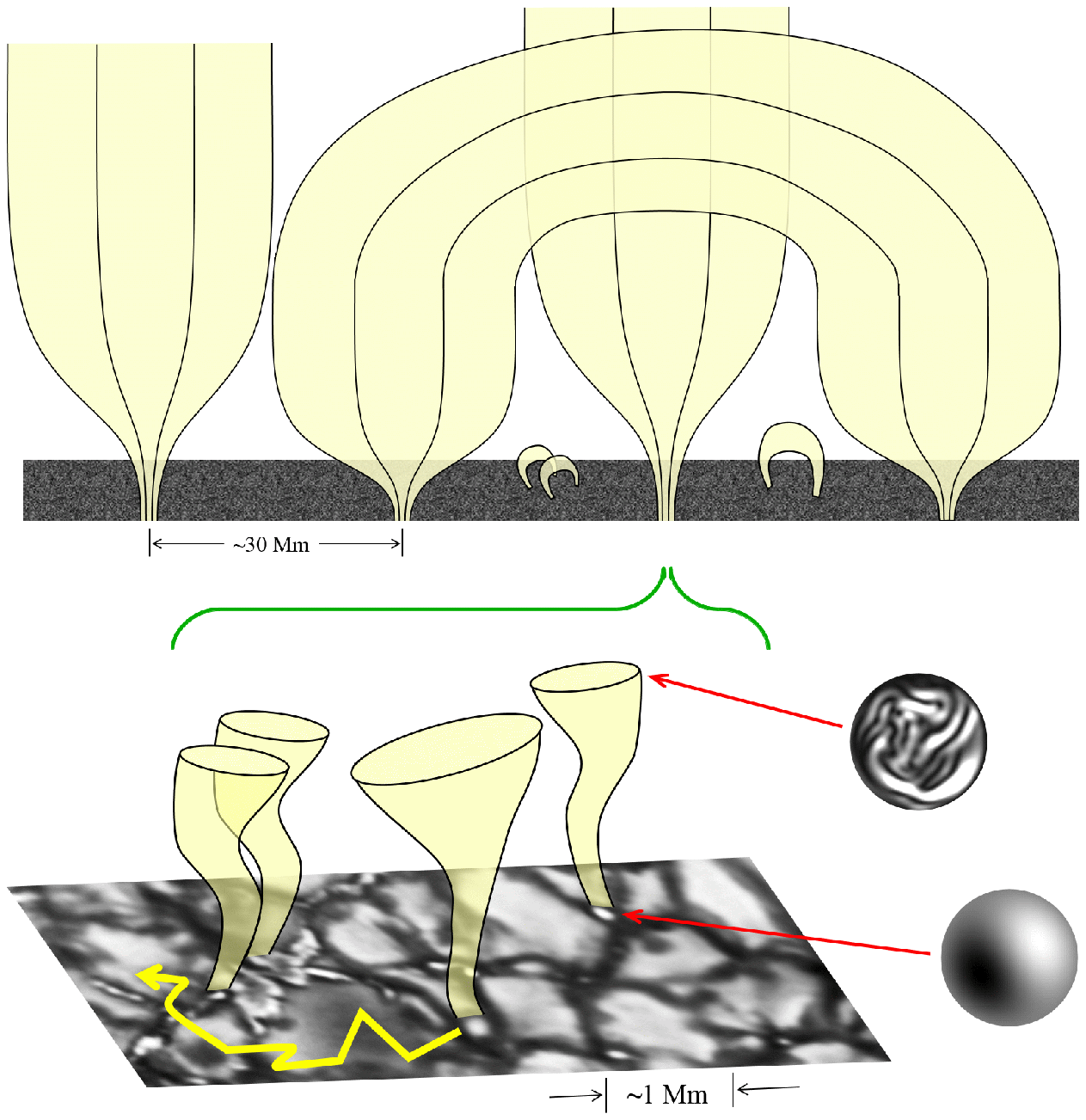}
\caption{Cartoon representation of coronal magnetic field expansion,
inspired by similar figures from \citet{Ga76}, \citet{Pe01},
\citet{CvB05}, and \citet{Ti10}.
{\em Top:} On supergranular scales the field is concentrated most
strongly in the bright chromospheric network.
Flux tubes expand laterally with increasing height to form a
funnel/canopy structure.
{\em Bottom:} On smaller scales, the field is concentrated into
intergranular flux tubes that are jostled by convection and
experience MHD turbulence.
Granulation image adapted from \citet{Ab11}, and turbulent
flux-tube cross sections adapted from \citet{AT13}.}
\label{fig01}
\end{figure}

Figure \ref{fig01} is a simplified illustration of how we believe the
complex multi-scale magnetic field is organized in regions that feed
the solar wind.
The most fundamental building blocks appear to be
the {\em G-band bright points,} i.e., thin flux tubes with
field strengths of 1000--2000 G and diameters of 50--200 km that are
herded into the dark downflow lanes between the granulation cells
\citep{BT01}.
These flux tubes appear to be oriented in a nearly vertical direction,
and they undergo cross-sectional expansion as the field strength
decreases with increasing height.
Their filling factor (i.e., the fraction of the solar surface they cover)
grows from values less than 0.1--1\% in the photosphere to a near
volume-filling unity in the low corona.
These features seem to be the ultimate ``footpoints'' of nearly
all larger-scale coronal magnetic structures.
Thus, the nature of their magnetohydrodynamic (MHD) variability
is necessarily of prime interest when attempting to understand the
evolution and dynamics of the coronal magnetic field.

Intergranular bright-point flux tubes are jostled continuously by
the evolution of surrounding granules.
Isolated flux tubes undergo a random-walk type of
motion \citep{vB98}, and occasionally one sees discrete
jumps that may indicate the merging or fragmenting of unresolved
components \citep{Bg98}.
Some bright points execute classical Brownian (i.e., erratic and
diffusive) random-walk motions, but others appear to move in a more
vortex-like or rotational manner \citep[see, e.g.,][]{Ya14}.
In the so-called thin-tube limit \citep{Sp81} it is possible to
decompose most observed bright-point motions into a superposition of
transverse, incompressible kink modes.
\citet{CvB05} showed how these kink modes gradually change their
character to match the classical transverse Alfv\'{e}n wave when
the thin tubes become volume-filling at larger heights.
Inside the flux tubes, the upward energy flux of Alfv\'{e}n waves
has been inferred to be about $3 \times 10^{5}$ W~m$^{-2}$.
However, after taking into account the small filling factors, the
{\em surface-averaged} photospheric flux is only about
2000 W~m$^{-2}$ \citep{CvB05}.
Nevertheless, this appears to be more than sufficient to balance
radiative losses in open-field regions \citep[e.g.,][]{WN77}.

In addition to the above idea of magnetic flux tubes being buffeted
and shuffled by granular motions, these bundles of magnetic field are
also continuously emerging from below the photosphere.
Away from active regions, much of this emergence takes the form of
bipolar ``ephemeral regions'' \citep{HM73,Hg08}.
The individual poles of these regions tend to be advected to the
edges of supergranular cells, where they coalesce to form the
chromospheric network.
Along the way, a combination of diffusive shuffling, tangling, and
magnetic reconnection with neighboring regions helps to maintain the
system in a dynamical steady state \citep[e.g.,][]{Sj97}.
We discuss the energetic consequences of this emergence and
cancellation for coronal heating in Section \ref{sec:rlo}.

There is some debate concerning the turbulent nature of the
photospheric magnetic elements.
High-resolution movies of individual granules often convey the
appearance of quite ``laminar'' overturning motions.
The kinetic energy spectrum of horizontal bright-point motions is
continuous in frequency space \citep[e.g.,][]{CvB05,Ct12}, but it
does not necessarily show the power-law behavior expected from an
active cascade.
Time-dependent MHD simulations of the chromosphere and corona
(see Section \ref{sec:wtd} below) typically do not need there to be
fully turbulent motions injected at the lower boundary, since a
rapid cascade develops naturally above the photosphere.
However, it is difficult to avoid the fact that the deep convection
zone---host to the highly nonlinear solar dynamo---must be strongly
turbulent \citep{Mm05}.
Observations of the spatial diffusion of magnetic elements in the
photosphere also show statistical properties similar to those seen
in simulations of turbulent diffusion \citep{Pt01,Ab11}.
Further elucidation will likely come from the next generation of
high-resolution solar observations, such as the 
Daniel K.\  Inouye Solar Telescope \citep[DKIST; see][]{Ri12}
and the Coronal Solar Magnetism Observatory 
\citep[COSMO; see][]{To13}.

\section{Waves and turbulence in the open-field corona}
\label{sec:wtd}

The idea that the solar atmosphere is heated by the dissipation of
upward-propagating waves has been studied for more than a half
century \citep[e.g.,][]{Os61}.
This remains a compelling explanation---especially for the
open-field regions that connect to the solar wind---because the
energy responsible for coronal heating must somehow propagate
up to the heights where the heat is deposited into the plasma.
{\em In~situ} measurements have shown that the fast
solar wind contains a dominant population of Alfv\'{e}n waves
\citep{BD71} that propagate mostly away from the Sun.
However, there has been uncertainty about whether Alfv\'{e}n waves
generated in the photosphere can damp out rapidly enough in the
low corona to provide the required heating \citep[see, e.g.,][]{P91}.

The last few decades have seen a gradual refinement in
models that make use of waves and turbulence to power the
solar wind \citep{Ho86,Ve91,Mt99}.
The convection-driven jostling of flux-tube footpoints
(see Section \ref{sec:photo}) may give rise to a range of MHD
wave modes, but it is likely that the incompressible Alfv\'{e}n
mode undergoes the least amount of viscous and conductive damping
to become the dominant wave type in the upper chromosphere and
low corona.
Alfv\'{e}n waves that propagate along a flux tube with a radially
varying Alfv\'{e}n speed undergo partial reflection back
toward the Sun \citep{HO80,Ve93}.
The existence of counter-propagating wave packets---even if only
a small fraction of the energy is coming back down---allows a
nonlinear MHD turbulent cascade to develop
\citep{Ir63,Kr65,Db80,Ve89}.
The steady-state cascade rate of strong, imbalanced, and
anisotropic MHD turbulence determines the rate of dissipation,
and thus the rate of coronal heating
\citep[see also][]{Ho95,Lw07,Ch09,Ou15,Co15}.

\citet{CvB07} described a project to build realistic, time-steady,
and one-fluid models of wave/turbulence-driven coronal heating in
open magnetic flux tubes.
In this numerical code, called ZEPHYR, the rate of partial wave
reflection is computed from linearized non-WKB transport equations.
The rate of wave dissipation (and plasma heating) is computed
from expressions derived from a wide range of models of
strong Alfv\'{e}nic cascade with variable cross helicity
\citep[see, e.g.,][]{Dm02}.
Once the turbulence physics is specified, the only freely
adjustable parameters in ZEPHYR are the photospheric lower boundary
conditions and the radial dependence of the flux tube's magnetic field.
\citet{CvB07} found that a fixed choice for the photospheric
wave properties yielded a realistic range of slow and fast solar
wind conditions.
More recently, this model has been extended to include a broader
range of magnetic configurations \citep{CvB13,WC14} in order
to explore how the properties of the solar wind depend
on the corona's ``superradial'' flux-tube expansion
\citep[see also][]{Ch11,Li14}.
Also, \citet{Cr14b} showed that wave/turbulence models can
reproduce {\em in~situ} ion charge states quite well if electron
heat conduction in the low corona is treated realistically (i.e.,
following the non-classical development of weakly collisional
suprathermal tails).

Despite the successes of time-steady models, they do not
self-consistently simulate the actual process of MHD turbulent
cascade at the heart of the proposed coronal heating scenario.
Thus, fully time-dependent and three-dimensional models of
reduced MHD turbulence were developed by \citet{vB11}.
This simulation code, called BRAID, follows the development of
turbulence in a flux tube rooted in the photosphere and driven
by slow laminar {\em internal} motions.\footnote{%
Although we focus on the insights gained from the BRAID code,
we also note that many other high-quality numerical simulations of
coronal turbulence have been performed
\citep[see, e.g.,][]{Ei96,Ni04,BV07,Rp07,Db12,PC13}.
This is a vibrant community, and the diversity of techniques and
perspectives has surely helped us all to better understand this
complex system.}
Turbulence develops rapidly in the chromosphere and is transmitted
into the corona (see cross sections shown in Figure \ref{fig01})
despite the large degree of reflection encountered at the sharp
transition region.
The time-dependent models constructed so far
\citep[see also][]{AT13} have largely validated the
phenomenological cascade rate prescriptions used by ZEPHYR.
\citet{vB14} used the BRAID code to conclude that a family of
quasi-static ``direct current'' (DC) models of coronal heating
should be ruled out.
They found that waves and turbulence can give rise to a
strongly dynamical type of twisting/braiding of coronal field lines,
and this appears to be consistent with recent ultra-high-resolution
imagery \citep{Ci13}.

\begin{figure}[!t]
\centering\includegraphics[width=5.3in]{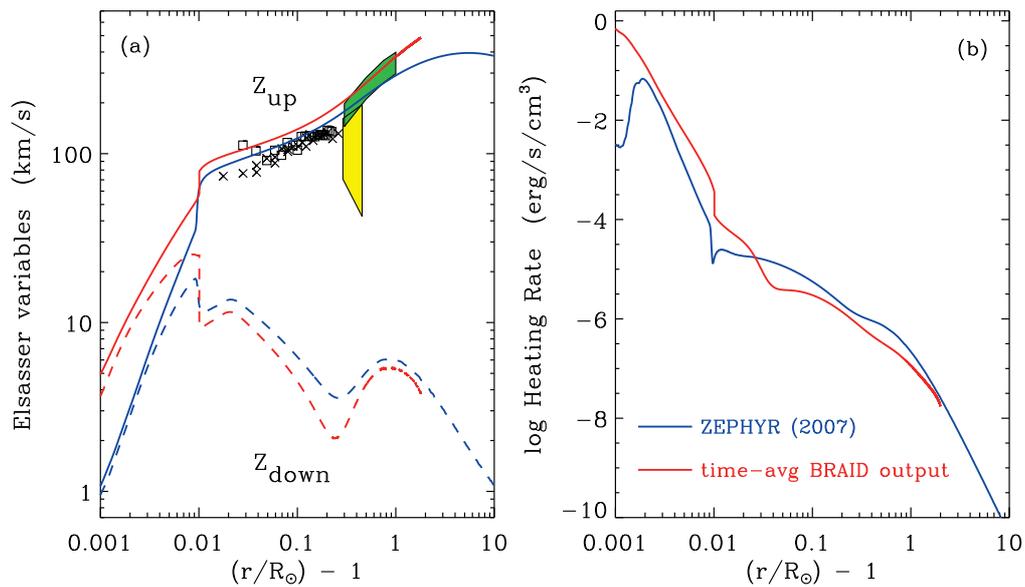}
\caption{Comparison of the height variation of
(a) Alfv\'{e}n wave velocity amplitudes, expressed as Elsasser
variables of upward $Z_{\rm up}$ and downward $Z_{\rm down}$
wave energy, and (b) the turbulent heating rate, from BRAID
(red curves) and ZEPHYR (blue curves).
Also shown are observational data for Alfv\'{e}n wave amplitudes,
converted into Elsasser variables under the assumption that
nearly all the energy is in $Z_{\rm up}$:
black X's \citep{Bn98}, black squares \citep{LC09},
green region \citep{Es99}, yellow region \citep{HS13}.}
\label{fig02}
\end{figure}

Figure~\ref{fig02} shows a preliminary comparison between the
results of ZEPHYR and BRAID for a model of a polar coronal hole.
The strongly fluctuating plasma parameters computed by BRAID were
averaged in time and truncated at a height of 2 $R_{\odot}$ above
the photosphere (above which it was assumed no incoming waves
enter the system).
There are similarities and differences between the two models
that still need to be digested and understood.
Note from Figure \ref{fig02}(a) that there is some tension
between different observational determinations of the Alfv\'{e}n
wave velocity amplitude at heights of 0.3--0.6 $R_{\odot}$ above
the photosphere.
Both \citet{Es99} and \citet{HS13} analyzed the nonthermal
broadening of ultraviolet emission lines observed above the
solar limb and interpreted them in terms of unresolved
Doppler ``sloshing'' due to Alfv\'{e}n waves.
However, there is not enough information in the data to fully
specify the full set of ion temperatures and wave amplitudes
for any given collection of emission lines.
These studies used different sets of assumptions---and different
data analysis techniques---to estimate the wave amplitudes.
It is still an open question whether the strong damping
inferred by \citet{HS13} actually exists.

\section{Magnetic reconnection in the open-field corona}
\label{sec:rlo}

It is clear from observations of the Sun's rapidly evolving
``magnetic carpet'' \citep{TS98} that much of the energy released
in the low corona is due to complex interactions between
neighboring loop-like concentrations of magnetic flux.
It is difficult to see any way around the idea that
{\em magnetic reconnection} is going on continuously in the corona
and converting magnetic free energy into heat.
Additionally, many of the open flux tubes that reach into the
heliosphere are rooted in the magnetic carpet, and their footpoints
are often in close proximity to the rapidly evolving closed loops.
Thus, it is natural to propose that some of the mass and energy
of the solar wind is injected from closed to open flux tubes by
reconnection \citep[see, e.g.,][]{Ax92,Fi99,Mo11,Rp12}.\footnote{%
Reconnection is also believed to be a dominant process in driving
coronal heating in the primarily {\em closed-field corona} of
active regions and the quiet Sun, but that topic is slightly
beyond the scope of this review
\citep[see, e.g,][]{P88,Pr02,L04,Kl06,PD12}.}

The idea of a reconnection-driven solar wind is relevant to a
discussion of coronal turbulence, since the two phenomena
(turbulence and reconnection) are nearly always seen in tandem.
On the one hand, the rate of driven reconnection may be
modulated---or even determined---by the properties of
``background'' turbulence \citep{Lz15}.
On the other hand, the dissipation of an MHD cascade is likely to
be governed by the behavior of spontaneously formed reconnection
regions on kinetic scales \citep[e.g.,][]{Os11,Mt15}.
The situation is complicated by the fact that the Poynting flux
associated with the emergence of closed-loop regions (which
presumably drive the reconnection) is seen to be roughly
$10^3$ W~m$^{-2}$ \citep{Fi99}, which is of the same order of
magnitude as the Alfv\'{e}n wave energy flux discussed above.
Thus, it would not be surprising if the corona were heated by an
inseparable combination of wave/turbulence dissipation and magnetic
reconnection.

Observations of coronal plumes and jets provide evidence for the
existence of discrete reconnection events in open-field regions
\citep{Sh07,Ti14},
However, these bright and narrow features are identifiable in images
because they appear to occupy a small fraction of the coronal
volume at any one time.
We thus need to determine how much of the solar wind's total mass
and energy is fed by these events.
This is related to finding a way to measure the fraction of the
corona's total ongoing reconnection that goes specifically into
opening up previously closed fields.
\citet{CvB10} studied these issues with Monte Carlo simulations of the
time-varying magnetic carpet and its connection to the large-scale
coronal field.
These models showed that reconnection and loop-opening processes on
supergranular scales may be responsible for the observed jets, but
probably not for the majority of ``bulk'' solar wind acceleration.
The \citet{CvB10} models were limited in scope because they approximated
the evolving corona as a succession of potential-field states, but
subsequent non-potential models \citep[e.g.,][]{Ct14} have found
similar results.

Recently there have been several other studies of how
magnetic reconnection may {\em indirectly} affect the energization
of solar wind plasma even if it does not provide the bulk of the
thermal energy.
Bursty reconnection events may be an additional source of MHD waves
in the corona \citep{Ho06,Ly14}.
Even if reconnection does not send a substantial energy flux into
the solar wind, it may help to ``fill in'' the
low-frequency part of the Alfv\'{e}nic fluctuation spectrum.
Periods longer than about 30 minutes are inferred to exist from the
{\em in~situ} data but are not found in models that rely
only on granular buffeting in the photosphere.
An alternate set of ideas was suggested by \citet{An11}, who
highlighted the existence of topologically complex
corridors of open fields rooted in distorted
{\em quasi-separatrix layers.}
These regions trace out a convoluted boundary between open and
closed field regions, and the forcing of the magnetic carpet
leads to its continual rearrangement.
Reconnection-driven loop opening is likely to occur in these regions,
and \citet{An11} suggested this ``separatrix web'' is a major driver
of the chaotic and dense slow wind.

\section{Physical insights from ultraviolet spectroscopy}
\label{sec:uvcs}

Many important clues about the physical processes responsible for
heating and accelerating the solar wind have come from ultraviolet
spectroscopy.
Because of the rapid decrease in density with increasing height,
Coulomb collisions become infrequent in the extended corona.\footnote{%
The term ``extended corona'' is meant to describe heights above
a few tenths of a solar radius above the photosphere, at which
coronagraphic occultation of the bright disk becomes necessary.
This region also tends to coincide with the main
``acceleration region'' of the solar wind.}
Thus, at a sufficiently large height,
each ion species may end up with a different temperature and
flow speed, and may even exhibit its own unique type of departure
from an isotropic Maxwellian velocity distribution.
This makes the collisionless outer corona a key place to
discriminate between different theories, since each process is likely
to suggest different rates of energization as a function of ion
charge and mass.
This section provides a brief ``top-ten list'' of results
from the Ultraviolet Coronagraph Spectrometer
\citep[UVCS; see][]{Ko97,Ko06}
that have helped to elucidate the properties of waves and
turbulence in the extended corona.
UVCS operated aboard the {\em Solar and Heliospheric Observatory}
({\em{SOHO}}) spacecraft from its launch in December 1995 until
the instrument's shutdown in January 2013.
Other useful reviews of the UVCS mission include those by
\citet{No97}, \citet{Hu10}, and \citet{An12}.

\subsection{Fast wind from coronal holes}

\begin{enumerate}

\item[1.]
UVCS observations of the bright O~VI 103.2, 103.7 nm emission lines
indicated that the associated O$^{+5}$ ions are much more strongly
heated than protons in polar coronal holes.
Detailed analysis of the line profile shapes and intensity ratios
point to perpendicular temperatures
$T_{\perp} \approx 2 \times 10^{8}$~K, roughly two orders of magnitude
hotter than the protons, and large temperature anisotropy ratios
$T_{\perp} / T_{\parallel}$ possibly greater than 10
\citep{Ko97,Cr99,Crao}.
Measured kinetic temperatures of both O$^{+5}$ and Mg$^{+9}$ ions
are greater than ``mass-proportional'' when compared with
protons, with $T_{i}/T_{p} > m_{i}/m_{p}$.
These conditions are reminiscent of the predicted properties of 
{\em ion cyclotron resonance} in a collisionless plasma.
Thus, the UVCS results helped to drive a resurgence of interest in
this mechanism in the fast wind
\citep[see, e.g.,][]{HI02,Ma06,Cr14a,Ga15}.

\item[2.]
UVCS line profile measurements in coronal holes also led to the
conclusion that the fast solar wind is accelerated closer to the
Sun than was believed in prior decades.
This kind of determination is made possible by the fact that
resonantly excited lines undergo ``Doppler dimming,'' i.e., they are
sensitive to motions transverse to the line of sight and thus
mostly parallel to the off-limb magnetic field.
\citet{St93} used sounding rocket data to infer a supersonic speed
($v \approx 200$ km s$^{-1}$ at $r \approx 2 \, R_{\odot}$)
for protons in a coronal hole.
UVCS then made it possible to determine that the outflow velocity
for O$^{+5}$ ions grows even faster than the outflow velocity for
protons.
By $r \approx 3$ to 3.5 $R_{\odot}$, the heavy ions are flowing
faster than the ``bulk'' solar wind by as much as a factor of two
\citep{Ko97,Cr99,Crao,An00}.

\item[3.]
In polar coronal holes, the densest concentrations of bright plumes
are seen to have higher densities, lower temperatures, and lower
outflow speeds than the dimmer ``interplume'' regions
\citep[e.g,.][]{Ko97,Gi00,Wi11}.
Many impulsive polar jets have similar properties as the longer-lived
plumes \citep{Db02}.
However, there may be observational selection effects at work, since
the UVCS counterparts to jets first identified by {\em Hinode} at
X-ray wavelengths seem to show hotter protons than in the surrounding
regions \citep{Mi07}.
Even if the coronal response at large heights is variable, the
ultraviolet observations of jets appear to be consistent with
models \citep{Wa94} in which a short burst of heating occurs at the
base then fades away with time.
These features may be a prime example of regions in which both
reconnection and turbulence are acting together.

\item[4.]
UVCS measured the properties of other types of coronal holes
that appear and disappear at various times throughout the 11-year
solar cycle.
\citet{Mi01} found that large {\em equatorial holes} undergo more
gradual solar wind acceleration than the polar holes, but they
both eventually reach high speeds ($v > 600$ km s$^{-1}$) at 1~AU.
Statistical studies of coronal holes at all latitudes show a
strong correlation between O$^{+5}$ temperatures and outflow speeds
\citep{Mi04,Mi10}, indicating that the preferential perpendicular heating
is tightly coupled with strong differential acceleration of heavy ions.

\end{enumerate}

\subsection{Slow wind from streamers}

\begin{enumerate}

\item[5.]
Much of the low-speed solar wind appears to be associated with
bright {\em helmet streamers.}
White-light images show most streamers having a closed-field
``core'' surrounded by ``legs'' that are open to the solar wind,
often topped by a converged radial ``stalk.''
UVCS Doppler dimming measurements at solar minimum \citep{St02}
revealed the legs of large equatorial streamers to be a primary
site of slow-wind-like outflow, whereas their large central cores
did not show signs of bulk acceleration.
Above the largest height of closed-field loops, O$^{+5}$ ions
in streamer stalks were seen to have similar preferential heating
characteristics as their cousins in coronal holes \citep{Fz03}.
Also, \citet{St12} found some key correlations in streamers between
UVCS plasma parameters and the relative rates of magnetic
flux-tube expansion, which \citet{WS90} suggested could be a
controlling factor in solar wind acceleration.
These correlations must be confronted and explained by any
successful model; for an initial comparison of this type, see
\citet{Ab10}.

\item[6.]
\citet{Ra97} found that the elemental abundances of heavy ions in
streamer legs match those seen {\em in~situ} in the slow wind.
However, in the closed-field cores the abundances are depleted by
more than an order of magnitude \citep{Ra99,VR05}.
This has been cited as evidence for gravitational settling, which
begins to occur when Coulomb collisions are not fast enough to keep
the ions mixed with the dominant hydrogen plasma.
Because the settling effect alone would have depleted the abundances
by much more than is seen, however, \citet{Ra99} suggested that
streamers are likely to host some additional process that
continuously mixes ``fresh'' plasma into the streamer core.
Coulomb drag and interchange reconnection may contribute to the
cross-field mixing, but turbulence is also a strong candidate.
Ubiquitous nonthermal line widths of order 30 km s$^{-1}$ in the low
corona point to Alfv\'{e}n-like fluctuations as either the means
of additional spatial transport, or as a source of turbulent pressure
that could support the ions in a non-hydrostatic equilibrium.

\item[7.]
UVCS measurements of non-equatorial streamers have shown significant
variation in their plasma properties.
Some mid-latitude streamers associated with active regions were seen
to have lower ion temperatures, but higher electron temperatures,
than large ``quiescent'' solar-minimum streamers
\citep{Fz99,Pa00,Ve05}.
The reasons for these differences may be explainable with a newer
classification that is based on magnetic polarity:
classical streamers with opposite-polarity footpoints (topped by
null-point cusps) appear to have distinct properties from the
so-called {\em pseudostreamers} with like-polarity footpoints
\citep[see, e.g.,][]{Wa12}.
Efforts are underway to identify the remote-sensing and {\em in~situ}
data that best distinguishes these structures from one another and
identifies the relevant physical properties at work \citep{Mi14}.

\end{enumerate}

\subsection{Coronal mass ejections}

\begin{enumerate}

\item[8.]
Coronal mass ejections (CMEs) are magnetically driven eruptions
that are believed to involve the expansion of twisted ``flux ropes''
into the heliosphere.
UVCS spectroscopy provided the first real diagnostics of the physical
conditions in CME plasmas as they accelerate through
the corona.
Specifically, UVCS data helped to determine that CMEs must undergo
substantial heating \citep{Ra08,Mu11}, and in some cases the input
thermal energy even exceeds the bulk kinetic energy.
It is possible that similar kinds of turbulent cascade and
dissipation occur in both CMEs and the background solar wind
\citep{Lu06}, or that these highly twisted, non-potential structures
may dissipate energy via turbulent hyperdiffusion \citep{vB08}.
In any case, measurements of CME total energy budgets are key
constraints to determining the dominant processes at work in these
events.

\item[9.]
Many models of CME formation predict the development of an
elongated {\em current sheet} trailing behind the flux rope
\citep{Li08}.
When these reconnecting features exist, there may be significant
mass and energy injection into the CME's outer sheath.
UVCS observed these current sheets in the emission of high-ionization
species such as Fe$^{+16}$, Fe$^{+17}$, and Ca$^{+13}$
\citep{Ci02,Ra08,Ko10}.
In some cases, the width of the current sheet was found to be
consistent with an anomalously large resistivity
\citep[e.g.,][]{Bp08,Li09}, which may also be consistent with
theories of turbulent reconnection \citep{Lz15}.

\item[10.]
In a few events, UVCS detected the presence of a supra-Alfv\'{e}nic
shock in front of a fast CME \citep[e.g.,][]{Ra00,Ma02}.
Independent measurements of the shock speed and electron density
(from Type II radio bursts) allow the coronal Alfv\'{e}n speed,
shock Mach number, and compression ratio to be determined.
Also, the rapid broadening of some emission lines seen by UVCS
indicated preferential ion heating and relatively weak electron
heating, which has been predicted in some kinetic models of
collisionless shocks \citep{LW00}.

\end{enumerate}

\section{Conclusions}
\label{sec:conc}

The goal of this paper has been to review the progress made in
recent years in viewing the processes of coronal heating and
solar wind acceleration through the lens of MHD turbulence.
There are viable models that can explain the energization of the
open-field plasma as a result of the dissipation of
propagating waves and turbulent eddies.
There are also (somewhat less developed, but still plausible)
models that invoke magnetic reconnection and the opening up of
closed loops into the heliosphere.
Despite several decades of useful observational constraints,
we still have not yet pinpointed what specific measurements will
convincingly identify the dominant physical processes.
Nevertheless, it is clear that telescopes have not yet achieved
the necessary spatial and time resolution to see how the most
fundamental structures (e.g., flux tubes, turbulent wave packets)
exchange energy in the solar atmosphere.
Also, we realize that there is a pressing need to bridge the
``field-of-view gap'' between the historically separated domains
of remote sensing and {\em in~situ} exploration.
Inner heliospheric missions like {\em Solar Probe Plus}
\citep{Fx13} and {\em Solar Orbiter} \citep{Mu13} will lead this
charge in the coming decade \citep[see also][]{Go15}.
We also hope that next-generation ultraviolet coronagraph
spectrometers \citep[e.g.,][]{Ko08} may be developed in order to
follow up on the successes of UVCS and extend the remote-sensing
field of view to larger heights.

From a theoretical viewpoint,
one physical process that deserves additional attention
is {\em compressibility.}
Many of the models discussed above have been limited to the
mainly Alfv\'{e}nic, low plasma~$\beta$ limit in which density
fluctuations are either unimportant or completely ``passive.''
However, off-limb observations show that compressive fluctuations
propagate up in open-field regions, along with Alfv\'{e}n waves
\citep[e.g.,][]{KP12,Th13}.
The passage of Comet Lovejoy through the low corona
revealed the existence of sharp density striations \citep{Ra14},
and the relevant cross-field length scales are of the same order of
magnitude as the expected turbulence correlation length.
There have also been time-dependent MHD simulations in one
\citep{SI06} and two \citep{MS14} dimensions that show Alfv\'{e}n waves
being converted efficiently into fast or slow compressive modes.
Although the restricted dimensionality of these models is likely to be
preventing the most efficient mode of reduced-MHD cascade from
occurring \citep[see, e.g.,][]{Hw15},
there may be regions of the corona that do support this kind of
mode conversion.

Lastly, we believe it is worthwhile to summarize how an
understanding of turbulent physics in the corona is important
in a broader context.
A clear practical benefit from a better model of solar wind acceleration
is the possibility of improving long-term space weather forecasts
\citep[e.g.,][]{Ea08}.
A different technological connection---involving the phenomenon of
ion cyclotron resonance---may bear fruit in new designs for ion-based
rocket propulsion systems \citep{CD01}.
Of course, identifying and characterizing the physical processes at
work in the solar corona is useful in establishing a baseline of
knowledge that can be applied to other stars and more distant
astrophysical systems.
For example, plasma heating from an MHD turbulent cascade has been
suggested as being relevant to models of
the interstellar medium \citep{Hv13},
T~Tauri stars \citep{Cr08},
exoplanet atmospheres \citep{Tn13}, and
accretion disks around compact objects \citep{Mv00}.

\section*{Acknowledgment}

The authors gratefully acknowledge Adriaan van Ballegooijen and
John Kohl for valuable contributions to this research.
SRC's work was supported by NASA grants {NNX\-10\-AC11G}
and {NNX\-14\-AG99G}, and NSF SHINE program grant AGS-1259519.
MPM's work was supported by NASA grants {NNX\-06\-AG95G},
{NNX\-09\-AH22G}, and {NNX\-10\-AQ58G}.
The {\em{SOHO}} mission is a project of international cooperation
between NASA and ESA.


\bibliographystyle{apalike}

\begin{thebibliography}{ }

\bibitem[Abbo et al.(2010)]{Ab10}
Abbo, L., Antonucci, E., Mikic, Z., et al. 2010,
``Characterization of the slow wind in the outer corona,''
{\em Adv.\  Space Res.,} 46, 1400.

\bibitem[Abramenko et al.(2011)]{Ab11}
Abramenko, V. I., Carbone, V., Yurchyshyn, V., et al. 2011,
``Turbulent diffusion in the photosphere as derived from photospheric
bright point motion,''
{\em Astrophys.\  J.,} 743, 133.

\bibitem[Antiochos et al.(2011)]{An11}
Antiochos, S. K., Miki\'{c}, Z., Titov, V., et al. 2011,
``A model for the sources of the slow solar wind,''
{\em Astrophys.\  J.,} 731, 112.

\bibitem[Antonucci et al.(2012)]{An12}
Antonucci, E., Abbo, L., and Telloni, D. 2012,
``UVCS Observations of Temperature and Velocity Profiles in Coronal Holes,''
{\em Space Sci.\  Rev.} 172, 5.

\bibitem[Antonucci et al.(2000)]{An00}
Antonucci, E., Dodero, M. A., and Giordano, S. 2000,
``Fast Solar Wind Velocity in a Polar Coronal Hole during Solar Minimum,''
{\em Solar Phys.,} 197, 115.

\bibitem[Asgari-Targhi et al.(2013)]{AT13}
Asgari-Targhi, M., van Ballegooijen, A. A., Cranmer, S. R., et al. 2013,
``The spatial and temporal dependence of coronal heating by Alfv\'{e}n
wave turbulence,''
{\em Astrophys.\  J.,} 773, 111.

\bibitem[Axford and McKenzie(1992)]{Ax92}
Axford, W. I., and McKenzie, J. F. 1992,
``The origin of high speed solar wind streams,''
in {\em Solar Wind Seven,} ed.\  E.\  Marsch and R.\  Schwenn
(New York: Pergamon), 1.

\bibitem[Banerjee et al.(1998)]{Bn98}
Banerjee, D., Teriaca, L., Doyle, J. G., and Wilhelm, K. 1998,
``Broadening of SI VIII lines observed in the solar polar coronal holes,''
{\em Astron.\  Astrophys.,} 339, 208.

\bibitem[Belcher and Davis(1971)]{BD71}
Belcher, J. W., and Davis, L., Jr. 1971,
``Large-amplitude Alfv\'{e}n waves in the interplanetary medium, 2,''
{\em J.\  Geophys.\  Res.,} 76, 3534.

\bibitem[Bemporad(2008)]{Bp08}
Bemporad, A. 2008,
``Spectroscopic detection of turbulence in post-CME current sheets,''
{\em Astrophys.\  J.,} 689, 572.

\bibitem[Berger et al.(1998)]{Bg98}
Berger, T. E., L\"{o}fdahl, M. G., Shine, R. S., et al. 1998,
``Measurements of solar magnetic element motion from high-resolution
filtergrams,''
{\em Astrophys.\  J.,} 495, 973.

\bibitem[Berger and Title(2001)]{BT01}
Berger, T. E., and Title, A. M. 2001,
``On the relation of G-band bright points to the photospheric magnetic field,''
{\em Astrophys.\  J.,} 553, 449.

\bibitem[Buchlin and Velli(2007)]{BV07}
Buchlin, E., and Velli, M. 2007,
``Shell Models of RMHD Turbulence and the Heating of Solar Coronal Loops,''
{\em Astrophys.\  J.,} 662, 701.

\bibitem[Chandran et al.(2011)]{Ch11}
Chandran, B. D. G., Dennis, T. J., Quataert, E., et al. 2011,
``Incorporating kinetic physics into a two-fluid solar-wind model with
temperature anisotropy and low-frequency Alfv\'{e}n-wave turbulence,''
{\em Astrophys.\  J.,} 743, 197.

\bibitem[Chandran et al.(2009)]{Ch09}
Chandran, B. D. G., Quataert, E., Howes, G. G., et al. 2009,
``The Turbulent Heating Rate in Strong Magnetohydrodynamic Turbulence
with Nonzero Cross Helicity,''
{\em Astrophys.\  J.,} 701, 652.

\bibitem[Chang D\'{\i}az(2001)]{CD01}
Chang D\'{\i}az, F. R. 2001,
``An overview of the VASIMR engine: High power space propulsion
with RF plasma generation and heating,''
in {\em Radio Frequency Power in Plasmas,}
AIP Conf.\  Proc.\  595, (New York: AIP Press), 3.

\bibitem[Chitta et al.(2014)]{Ct14}
Chitta, L. P., Kariyappa, R., van Ballegooijen, A. A., et al. 2014,
``Nonlinear force-free field modeling of the solar magnetic carpet
and comparison with SDO/HMI and Sunrise/IMaX observations,''
{\em Astrophys.\  J.,} 793, 112.

\bibitem[Chitta et al.(2012)]{Ct12}
Chitta, L. P., van Ballegooijen, A. A., Rouppe van der Voort, L.,
et al. 2012,
``Dynamics of the solar magnetic bright points derived from their
horizontal motions,''
{\em Astrophys.\  J.,} 752, 48.

\bibitem[Ciaravella et al.(2002)]{Ci02}
Ciaravella, A., Raymond, J. C., Li, J., et al. 2002,
``Elemental abundances and post-coronal mass ejection current sheet
in a very hot active region,''
{\em Astrophys.\  J.,} 575, 1116.

\bibitem[Cirtain et al.(2013)]{Ci13}
Cirtain, J. W., Golub, L., Winebarger, A. R., et al. 2013,
``Energy release in the solar corona from spatially resolved
magnetic braids,''
{\em Nature,} 493, 501.

\bibitem[Coburn et al.(2015)]{Co15}
Coburn, J. T., Forman, M. A., Smith, C. W., et al. 2015,
``Third-Moment Descriptions of the Interplanetary Turbulent Cascade,
Intermittency, and Back Transfer,''
{\em Phil.\  Trans.\  R.\  Soc.\  A,} this issue.

\bibitem[Cranmer(2008)]{Cr08}
Cranmer, S. R. 2008,
``Turbulence-driven polar winds from T Tauri stars
energized by magnetospheric accretion,''
{\em Astrophys.\  J.,} 689, 316.

\bibitem[Cranmer(2014a)]{Cr14a}
Cranmer, S. R. 2014a,
``Ensemble simulations of proton heating in the solar wind via
turbulence and ion cyclotron resonance,''
{\em Astrophys.\  J.\  Suppl.,} 213, 16.

\bibitem[Cranmer(2014b)]{Cr14b}
Cranmer, S. R. 2014b,
``Suprathermal electrons in the solar corona: Can nonlocal transport
explain heliospheric charge states?''
{\em Astrophys.\  J.,} 791, L31.

\bibitem[Cranmer et al.(1999)]{Cr99}
Cranmer, S. R., Kohl, J. L., Noci, G., et al. 1999,
``An empirical model of a polar coronal hole at solar minimum,''
{\em Astrophys.\  J.,} 511, 481.

\bibitem[Cranmer et al.(2008)]{Crao}
Cranmer, S. R., Panasyuk, A. V., and Kohl, J. L. 2008,
``Improved constraints on the preferential heating and acceleration
of oxygen ions in the extended solar corona,''
{\em Astrophys.\  J.,} 678, 1480.

\bibitem[Cranmer and van Ballegooijen(2005)]{CvB05}
Cranmer, S. R., and van Ballegooijen, A. A. 2005,
``On the generation, propagation, and reflection of Alfv\'{e}n waves
from the solar photosphere to the distant heliosphere,''
{\em Astrophys.\  J.\  Suppl.,} 156, 265.

\bibitem[Cranmer and van Ballegooijen(2010)]{CvB10}
Cranmer, S. R., and van Ballegooijen, A. A. 2010,
``Can the solar wind be driven by magnetic reconnection in the Sun's
magnetic carpet?''
{\em Astrophys.\  J.,} 720, 824.

\bibitem[Cranmer et al.(2007)]{CvB07}
Cranmer, S. R., van Ballegooijen, A. A., and Edgar, R. J. 2007,
``Self-consistent coronal heating and solar wind acceleration
from anisotropic magnetohydrodynamic turbulence,''
{\em Astrophys.\  J.\  Suppl.,} 171, 520.

\bibitem[Cranmer et al.(2013)]{CvB13}
Cranmer, S. R., van Ballegooijen, A. A., and Woolsey, L. N. 2013,
``Connecting the Sun's high-resolution magnetic carpet to
the turbulent heliosphere,''
{\em Astrophys.\  J.,} 767, 125.

\bibitem[Dahlburg et al.(2012)]{Db12}
Dahlburg, R. B., Einaudi, G., Rappazzo, A. F., and Velli, M. 2012,
``Turbulent coronal heating mechanisms: coupling of dynamics and
thermodynamics,''
{\em Astron.\  Astrophys.,} 544, L20.

\bibitem[Dmitruk et al.(2002)]{Dm02}
Dmitruk, P., Matthaeus, W. H., Milano, L. J., et al. 2002,
``Coronal heating distribution due to low-frequency, wave-driven
turbulence,''
{\em Astrophys.\  J.,} 575, 571.

\bibitem[Dobrowolny et al.(1980)]{Db80}
Dobrowolny, M. and Mangeney, A. and Veltri, P. 1980,
``Fully developed anisotropic hydromagnetic turbulence in
interplanetary space,''
{\em Phys.\  Rev.\  Lett.,} 45, 144.

\bibitem[Dobrzycka et al.(2002)]{Db02}
Dobrzycka, D., Cranmer, S. R., Raymond, J. C., et al. 2002,
``Polar coronal jets at solar minimum,''
{\em Astrophys.\  J.,} 565, 621.

\bibitem[Eastwood(2008)]{Ea08}
Eastwood, J. P. 2008, ``The science of space weather,''
{\em Phil.\  Trans.\  R.\  Soc.\  A,} 366, 4489.

\bibitem[Einaudi et al.(1996)]{Ei96}
Einaudi, G., Velli, M., Politano, H., et al. 1996,
``Energy Release in a Turbulent Corona,''
{\em Astrophys.\  J.,} 457, L113.

\bibitem[Esser et al.(1999)]{Es99}
Esser, R., Fineschi, S., Dobrzycka, D., et al. 1999,
``Plasma properties in coronal holes derived from measurements of
minor ion spectral lines and polarized white light intensity,''
{\em Astrophys.\  J.,} 510, L63.


\bibitem[Fisk et al.(1999)]{Fi99}
Fisk, L. A., Schwadron, N. A., and Zurbuchen, T. H. 1999,
``Acceleration of the fast solar wind by the emergence of new magnetic
flux,'' {\em J.\  Geophys.\  Res.,} 104, 19765.

\bibitem[Fox et al.(2013)]{Fx13}
Fox, N. J., Bale, S. D., Decker, R. B., et al. 2013,
``Solar Probe Plus: A NASA mission to touch the Sun,''
{\em Eos Trans. AGU,} Fall 2013 Meet. Suppl., abstract SM53A--2207.

\bibitem[Frazin et al.(1999)]{Fz99}
Frazin, R. A., Ciaravella, A., Dennis, E., et al. 1999,
``UVCS/SOHO ion kinetics in coronal streamers,''
{\em Space Sci.\  Rev.} 87, 189.

\bibitem[Frazin et al.(2003)]{Fz03}
Frazin, R. A., Cranmer, S. R., and Kohl, J. L. 2003,
``Empirically determined anisotropic velocity distributions and
outflows of O$^{5+}$ ions in a coronal streamer at solar minimum,''
{\em Astrophys.\  J.,} 597, 1145.

\bibitem[Gabriel(1976)]{Ga76}
Gabriel, A. H. 1967,
``A magnetic model of the solar transition region,''
{\em Phil.\  Trans.\  R.\  Soc.\  A,} 281, 339.

\bibitem[Gary(2015)]{Ga15}
Gary, S. P. 2015,
``Short-Wavelength Turbulence and Temperature Anisotropy
Instabilities: Recent Computational Progress,''
{\em Phil.\  Trans.\  R.\  Soc.\  A,} this issue.

\bibitem[Giordano et al.(2000)]{Gi00}
Giordano, S., Antonucci, E., Noci, G., et al. 2000,
``Identification of the coronal sources of the fast solar wind,''
{\em Astrophys.\  J.,} 531, L79.

\bibitem[Goldstein et al.(2015)]{Go15}
Goldstein, M. L., Wicks, R. T., Perri, S., and Sahraoui, F. 2015,
``Kinetic scale turbulence and dissipation in the solar wind:
Key observational results and future outlook,''
{\em Phil.\  Trans.\  R.\  Soc.\  A,} this issue.

\bibitem[Hagenaar et al.(2008)]{Hg08}
Hagenaar, H. J., De Rosa, M. L., and Schrijver, C. J. 2008,
``The Dependence of Ephemeral Region Emergence on Local Flux Imbalance,''
{\em Astrophys.\  J.,} 678, 541.

\bibitem[Hahn and Savin(2013)]{HS13}
Hahn, M., and Savin, D. W. 2013,
``Observational quantification of the energy dissipated by Alfv\'{e}n
waves in a polar coronal hole: Evidence that waves drive the fast
solar wind,''
{\em Astrophys.\  J.,} 776, 78.

\bibitem[Hanasoge and Sreenivasan(2014)]{HS14}
Hanasoge, S. M., and Sreenivasan, K. R. 2014,
``The quest to understand supergranulation and large-scale convection
in the Sun,'' {\em Solar Phys.,} 289, 3403.

\bibitem[Harvey and Martin(1973)]{HM73}
Harvey, K. L., and Martin, S. F. 1973,
``Ephemeral Active Regions,'' {\em Solar Phys.,} 32, 389.

\bibitem[Haverkorn and Spangler(2013)]{Hv13}
Haverkorn, M., and Spangler, S. R. 2013,
``Plasma diagnostics of the interstellar medium with radio astronomy,''
{\em Space Sci.\  Rev.} 178, 483.

\bibitem[Heinemann and Olbert(1980)]{HO80}
Heinemann, M., and Olbert, S. 1980,
``Non-WKB Alfv\'{e}n waves in the solar wind,''
{\em J.\  Geophys.\  Res.,} 85, 1311.

\bibitem[Hoeksema and Scherrer(1986)]{HS86}
Hoeksema, J. T., and Scherrer, P. H. 1986,
``An atlas of photospheric magnetic field observations and computed
coronal magnetic fields: 1976--1985,''
{\em Solar Phys.,} 105, 205.

\bibitem[Hollweg(1986)]{Ho86}
Hollweg, J. V. 1986,
``Transition region, corona, and solar wind in coronal holes,''
{\em J.\  Geophys.\  Res.,} 91, 4111.

\bibitem[Hollweg(2006)]{Ho06}
Hollweg, J. V. 2006, ``Drivers of the solar wind: Then and now,''
{\em Phil.\  Trans.\  R.\  Soc.\  A,} 364, 505.

\bibitem[Hollweg and Isenberg(2002)]{HI02}
Hollweg, J. V., and Isenberg, P. A. 2002,
``Generation of the fast solar wind: A review with emphasis on the
resonant cyclotron interaction,''
{\em J.\  Geophys.\  Res.,} 107, 1147.

\bibitem[Hossain et al.(1995)]{Ho95}
Hossain, M., Gray, P. C., Pontius, D. H., Jr., et al. 1995,
``Phenomenology for the decay of energy-containing eddies in
homogeneous MHD turbulence,''
{\em Phys.\  Fluids,} 7, 2886.

\bibitem[Howes(2015)]{Hw15}
Howes, G. G. 2015,
``A dynamical model of plasma turbulence in the solar wind,''
{\em Phil.\  Trans.\  R.\  Soc.\  A,} this issue.


\bibitem[Huber(2010)]{Hu10}
Huber, M. C. E. 2010,
``The Role of Harvard College Observatory and UVCS in the
Development of SOHO,''
in {\em SOHO-23: Understanding a Peculiar Solar Minimum,}
ASP Conf.\  Ser.\  428, 15.

\bibitem[Hufbauer(1991)]{Hf91}
Hufbauer, K. 1991,
{\em Exploring the Sun: Solar Science Since Galileo}
(Baltimore: Johns Hopkins University Press).

\bibitem[Iroshnikov(1963)]{Ir63}
Iroshnikov, P. S. 1963,
``Turbulence of a conducting fluid in a strong magnetic field,''
{\em Astron.\  Zhurnal,} 40, 742.

\bibitem[Klimchuk(2006)]{Kl06}
Klimchuk, J. A. 2006,
``On solving the coronal heating problem,''
{\em Solar Phys.,} 234, 41.


\bibitem[Ko et al.(2010)]{Ko10}
Ko, Y.-K., Raymond, J. C., Vr\v{s}nak, B., et al. 2010,
``Modeling UV and X-ray emission in a post-coronal mass ejection
current sheet,''
{\em Astrophys.\  J.,} 722, 625.

\bibitem[Kohl et al.(2008)]{Ko08}
Kohl, J. L., Jain, R., Cranmer, S. R., et al. 2008,
``Next generation UV coronagraph instrumentation for solar cycle 24,''
\mbox{J. Astrophys. Astron.,} 29, 321.

\bibitem[Kohl et al.(1997)]{Ko97}
Kohl, J. L., Noci, G., Antonucci, E., et al. 1997,
``First results from the SOHO Ultraviolet Coronagraph Spectrometer,''
{\em Solar Phys.,} 175, 613.

\bibitem[Kohl et al.(2006)]{Ko06}
Kohl, J. L., Noci, G., Cranmer, S. R., and Raymond, J. C. 2006,
``Ultraviolet spectroscopy of the extended solar corona,''
{\em Astron.\  Astrophys.\  Review,} 13, 31.

\bibitem[Kraichnan(1965)]{Kr65}
Kraichnan, R. H. 1965,
``Inertial-Range Spectrum of Hydromagnetic Turbulence,''
{\em Phys.\  Fluids,} 8, 1385.

\bibitem[Krishna Prasad et al.(2012)]{KP12}
Krishna Prasad, S., Banerjee, D., Van Doorsselaere, T., and
Singh, J. 2012, ``Omnipresent long-period intensity oscillations in
open coronal structures,''
{\em Astron.\  Astrophys.,} 546, A50.

\bibitem[Landi and Cranmer(2009)]{LC09}
Landi, E., and Cranmer, S. R. 2009,
``Ion Temperatures in the Low Solar Corona: Polar Coronal Holes at
Solar Minimum,''
{\em Astrophys.\  J.,} 691, 794.

\bibitem[Lazarian et al.(2015)]{Lz15}
Lazarian, A., Eyink, G., Vishniac, E., and Kowal, G. 2015,
``Turbulent reconnection and its implications,''
{\em Phil.\  Trans.\  R.\  Soc.\  A,} this issue.


\bibitem[Leighton et al.(1962)]{Le62}
Leighton, R. B., Noyes, R. W., and Simon, G. W. 1962,
``Velocity Fields in the Solar Atmosphere, I, Preliminary Report,''
{\em Astrophys.\  J.,} 135, 474.

\bibitem[Lee and Wu(2000)]{LW00}
Lee, L. C., and Wu, B. H. 2000,
``Heating and Acceleration of Protons and Minor Ions by Fast Shocks
in the Solar Corona,''
{\em Astrophys.\  J.,} 535, 1014.

\bibitem[Lin et al.(2008)]{Li08}
Lin, J., Cranmer, S. R., and Farrugia, C. J. 2008,
``Plasmoids in reconnecting current sheets: Solar and terrestrial
contexts compared,''
{\em J.\  Geophys.\  Res.,} 113, A11107.

\bibitem[Lin et al.(2009)]{Li09}
Lin, J., Li, J., Ko, Y.-K., et al. 2009,
``Investigation of thickness and electrical resistivity of the
current sheets in solar eruptions,''
{\em Astrophys.\  J.,} 693, 1666.

\bibitem[Lionello et al.(2014)]{Li14}
Lionello, R., Velli, M., Downs, C., et al. 2014,
``Validating a time-dependent turbulence-driven model of the solar wind,''
{\em Astrophys.\  J.,} 784, 120.

\bibitem[Lithwick et al.(2007)]{Lw07}
Lithwick, Y., Goldreich, P., and Sridhar, S. 2007,
``Imbalanced strong MHD turbulence,''
{\em Astrophys.\  J.,} 655, 269.

\bibitem[Liu et al.(2006)]{Lu06}
Liu, Y., Richardson, J. D., Belcher, J. W., et al. 2006,
``Thermodynamic structure of collision-dominated expanding
plasma: Heating of interplanetary coronal mass ejections,''
{\em J.\  Geophys.\  Res.,} 111, A01102.

\bibitem[Longcope(2004)]{L04}
Longcope, D. W. 2004,
``Quantifying magnetic reconnection and the heat it generates,''
in {\em SOHO-15: Coronal Heating,} ed. R. W. Walsh, J. Ireland,
D. Danesy, and B. Fleck (Noordwijk, ESA), ESA SP-575, 198.

\bibitem[Lynch et al.(2014)]{Ly14}
Lynch, B. J., Edmondson, J. K., and Li, Y. 2014,
``Interchange reconnection Alfv\'{e}n wave generation,''
{\em Solar Phys.,} 289, 3043.

\bibitem[Mancuso et al.(2002)]{Ma02}
Mancuso, S., Raymond, J. C., Kohl, J. L., et al. 2002,
``UVCS/SOHO observations of a CME-driven shock: Consequences on
ion heating mechanisms behind a coronal shock,''
{\em Astron.\  Astrophys.,} 383, 267.

\bibitem[Marsch(2006)]{Ma06}
Marsch, E. 2006,
``Kinetic physics of the solar corona and solar wind,''
{\em Living Rev.\  Solar Phys.,} 3, 1.

\bibitem[Matsumoto and Suzuki(2014)]{MS14}
Matsumoto, T., and Suzuki, T. K. 2014,
``Connecting the Sun and the solar wind: the self-consistent transition
of heating mechanisms,''
{\em Mon.\  Not.\  Roy.\  Astron.\  Soc.,} 440, 971.

\bibitem[Matthaeus et al.(2015)]{Mt15}
Matthaeus, W. H., Wan, M., Servidio, S., et al. 2015,
``Intermittency, Nonlinear Dynamics, and Dissipation in the Solar Wind
and Astrophysical Plasmas,''
{\em Phil.\  Trans.\  R.\  Soc.\  A,} this issue.

\bibitem[Matthaeus et al.(1999)]{Mt99}
Matthaeus, W. H., Zank, G. P., Oughton, S., et al. 1999,
``Coronal heating by magnetohydrodynamic turbulence driven by
reflected low-frequency waves,''
{\em Astrophys.\  J.,} 523, L93.

\bibitem[Medvedev(2000)]{Mv00}
Medvedev, M. V. 2000,
``Particle heating by nonlinear Alfv\'{e}nic turbulence in
advection-dominated accretion flows.''
{\em Astrophys.\  J.,} 541, 811.

\bibitem[Miesch(2005)]{Mm05}
Miesch, M. S. 2005,
``Large-Scale Dynamics of the Convection Zone and Tachocline,''
{\em Living Rev.\  Solar Phys.,} 2, 1.

\bibitem[Miralles et al.(2004)]{Mi04}
Miralles, M. P., Cranmer, S. R., and Kohl, J. L. 2004,
``Low-latitude coronal holes during solar maximum,''
{\em Adv.\  Space Res.,} 33, 696.

\bibitem[Miralles et al.(2001)]{Mi01}
Miralles, M. P., Cranmer, S. R., Panasyuk, A. V., et al. 2001,
``Comparison of Empirical Models for Polar and Equatorial Coronal Holes,''
{\em Astrophys.\  J.,} 549, L257.

\bibitem[Miralles et al.(2010)]{Mi10}
Miralles, M. P., Cranmer, S. R., Panasyuk, A. V., et al. 2010,
``The Tale of Two Minima and a Solar Cycle in Between: An Ongoing
Fast Solar Wind Investigation,''
in {\em SOHO-23: Understanding a Peculiar Solar Minimum,}
ASP Conf.\  Series 428 (San Francisco: Astron.\  Soc.\  Pacific), 237.

\bibitem[Miralles et al.(2007)]{Mi07}
Miralles, M. P., Cranmer, S. R., Raymond, J. C., et al. 2007,
``Polar coronal jets during the 2007 joint SOHO/Hinode campaigns,''
{\em Eos Trans. AGU,} 88 (52), Fall Meeting Suppl., abstract SH21B-02. 

\bibitem[Miralles et al.(2014)]{Mi14}
Miralles, M. P., Cranmer, S. R., and Stenborg, G. A. 2014,
``Untangling Coronal Streamers from Pseudostreamers,''
{\em Bull.\  Am.\  Astron.\  Soc.,} AAS meeting 224, abstract 323.56.

\bibitem[Moore et al.(2011)]{Mo11}
Moore, R. L., Sterling, A. C., Cirtain, J. W., et al. 2011,
``Solar X-Ray Jets, Type II Spicules, Granule-size Emerging
Bipoles, and the Genesis of the Heliosphere,''
{\em Astrophys.\  J.,} 731, L18.

\bibitem[M\"{u}ller et al.(2013)]{Mu13}
M\"{u}ller, D., Marsden, R. G., St.\  Cyr, O. C., et al. 2013,
``Solar Orbiter: Exploring the Sun-Heliosphere Connection,''
{\em Solar Phys.,} 285, 25.
 
\bibitem[Murphy et al.(2011)]{Mu11}
Murphy, N. A., Raymond, J. C., and Korreck, K. E. 2011,
``Plasma Heating During a Coronal Mass Ejection Observed By the
Solar and Heliospheric Observatory,''
{\em Astrophys.\  J.,} 735, 17.

\bibitem[Neugebauer and Snyder(1966)]{NS66}
Neugebauer, M., and Snyder, C. W. 1966,
``Mariner 2 Observations of the Solar Wind, 1, Average Properties,''
{\em J.\  Geophys.\  Res.,} 71, 4469.

\bibitem[Nigro et al.(2004)]{Ni04}
Nigro, G., Malara, F., Carbone, V., et al. 2004,
``Nanoflares and MHD Turbulence in Coronal Loops: A Hybrid Shell Model,''
{\em Phys.\  Rev.\  Lett.,} 92, 194501.

\bibitem[Noci et al.(1997)]{No97}
Noci, G., Kohl, J. L., Antonucci, E., et al. 1997,
``First Results from UVCS/SOHO,''
{\em Adv.\  Space Res.,} 20, 2219.

\bibitem[Osman et al.(2011)]{Os11}
Osman, K. T., Matthaeus, W. H., Greco, A., et al. 2011,
``Evidence for inhomogeneous heating in the solar wind,''
{\em Astrophys.\  J.,} 727, L11.

\bibitem[Osterbrock(1961)]{Os61}
Osterbrock, D. E. 1961,
``The Heating of the Solar Chromosphere, Plages, and Corona by
Magnetohydrodynamic Waves,''
{\em Astrophys.\  J.,} 134, 347.

\bibitem[Oughton et al.(2015)]{Ou15}
Oughton, S., Matthaeus, W. H., Wan, M., and Osman, K. T. 2015,
``Anisotropy in solar wind plasma turbulence,''
{\em Phil.\  Trans.\  R.\  Soc.\  A,} this issue.

\bibitem[Parenti et al.(2000)]{Pa00}
Parenti, S., Bromage, B. J. I., Poletto, G., et al. 2000,
``Characteristics of solar coronal streamers: Element abundance,
temperature and density from coordinated CDS and UVCS SOHO observations,''
{\em Astron.\  Astrophys.,} 363, 800.

\bibitem[Parker(1958)]{P58}
Parker, E. N. 1958,
``Dynamics of the interplanetary gas and magnetic fields,''
{\em Astrophys.\  J.,} 128, 664.

\bibitem[Parker(1988)]{P88}
Parker, E. N. 1988,
``Nanoflares and the solar X-ray corona,''
{\em Astrophys.\  J.,} 330, 474.

\bibitem[Parker(1991)]{P91}
Parker, E. N. 1991,
``Heating solar coronal holes,''
{\em Astrophys.\  J.,} 372, 719.

\bibitem[Parker(2001)]{P01}
Parker, E. N. 2001,
``A history of early work on the heliospheric magnetic field,''
{\em J.\  Geophys.\  Res.,} 106, 15797.

\bibitem[Parnell and De Moortel(2012)]{PD12}
Parnell, C. E., and De Moortel, I. 2012,
``A contemporary view of coronal heating,''
{\em Phil.\  Trans.\  R.\  Soc.\  A,} 370, 3217.

\bibitem[Perez and Chandran(2013)]{PC13}
Perez, J. C., and Chandran, B. D. G. 2013,
``Direct numerical simulations of reflection-driven, reduced
magnetohydrodynamic turbulence from the Sun to the Alfv\'{e}n
critical point,''
{\em Astrophys.\  J.,} 776, 124.

\bibitem[Peter(2001)]{Pe01}
Peter, H. 2001,
``On the nature of the transition region from the chromosphere to
the corona of the Sun,''
{\em Astron.\  Astrophys.,} 374, 1108.

\bibitem[Petrovay(2001)]{Pt01}
Petrovay, K. 2001, ``Turbulence in the solar photosphere,''
{\em Space Sci.\  Rev.} 95, 9.

\bibitem[Priest et al.(2002)]{Pr02}
Priest, E. R., Heyvaerts, J. F., and Title, A. M. 2002,
``A flux-tube tectonics model for solar coronal heating driven by the
magnetic carpet,''
{\em Astrophys.\  J.,} 576, 533.

\bibitem[Rappazzo et al.(2012)]{Rp12}
Rappazzo, A. F., Matthaeus, W. H., Ruffolo, D., et al. 2012,
``Interchange reconnection in a turbulent corona,''
{\em Astrophys.\  J.,} 758, L14.

\bibitem[Rappazzo et al.(2007)]{Rp07}
Rappazzo, A. F., Velli, M., Einaudi, G., et al. 2007,
``Coronal Heating, Weak MHD Turbulence, and Scaling Laws,''
{\em Astrophys.\  J.,} 657, L47.

\bibitem[Rast(2003)]{Ra03}
Rast, M. P. 2003,
``The scales of granulation, mesogranulation, and supergranulation,''
{\em Astrophys.\  J.,} 597, 1200.

\bibitem[Raymond(1999)]{Ra99}
Raymond, J. C. 1999,
``Composition variations in the solar corona and solar wind,''
{\em Space Sci.\  Rev.} 87, 55.

\bibitem[Raymond(2008)]{Ra08}
Raymond, J. C. 2008,
``UV diagnostics for the energy budget of flares and CMEs,''
{\em J.\  Astrophys.\  Astron.,} 29, 187.

\bibitem[Raymond et al.(1997)]{Ra97}
Raymond, J. C., Kohl, J. L., Noci, G., et al. 1997,
``Composition of coronal streamers from the SOHO Ultraviolet
Coronagraph Spectrometer,''
{\em Solar Phys.,} 175, 645.

\bibitem[Raymond et al.(2014)]{Ra14}
Raymond, J. C., McCauley, P. I., Cranmer, S. R., and Downs, C. 2014,
``The Solar Corona as Probed by Comet Lovejoy (C/2011 W3),''
{\em Astrophys.\  J.,} 788, 152.

\bibitem[Raymond et al.(2000)]{Ra00}
Raymond, J. C., Thompson, B. J., St.~Cyr, O. C., et al. 2000,
``SOHO and radio observations of a CME shock wave,''
{\em Geophys.\  Res.\  Lett.,} 27, 1439.

\bibitem[Rimmele et al.(2012)]{Ri12}
Rimmele, T. R., Keil, S., McMullin, J., et al. 2012,
``Construction of the Advanced Technology Solar Telescope,''
in {\em The Second ATST--EAST Meeting: Magnetic Fields from the
Photosphere to the Corona,} ASP Conf.\  Series 463
(San Francisco: Astron.\  Soc.\  Pacific), 377.

\bibitem[Schrijver et al.(1997)]{Sj97}
Schrijver, C. J., Title, A. M., van Ballegooijen, A. A., et al. 1997,
``Sustaining the Quiet Photospheric Network: The Balance of Flux
Emergence, Fragmentation, Merging, and Cancellation,''
{\em Astrophys.\  J.,} 487, 424.

\bibitem[Shibata et al.(2007)]{Sh07}
Shibata, K., Nakamura, T., Matsumoto, T., et al. 2007,
``Chromospheric anemone jets as evidence of ubiquitous reconnection,''
{\em Science,} 318, 1591. 

\bibitem[Spruit(1981)]{Sp81}
Spruit, H. C. 1981, ``Motion of magnetic flux tubes in the solar
convection zone and chromosphere,''
{\em Astron.\  Astrophys.,} 98, 155.

\bibitem[Strachan et al.(1993)]{St93}
Strachan, L., Kohl, J. L., Weiser, H., et al. 1993,
``A Doppler dimming determination of coronal outflow velocity,''
{\em Astrophys.\  J.,} 412, 410.

\bibitem[Strachan et al.(2012)]{St12}
Strachan, L., Panasyuk, A. V., Kohl, J. L., et al. 2012,
``The Evolution of Plasma Parameters on a Coronal Source Surface at
2.3 $R_{\odot}$ during Solar Minimum,''
{\em Astrophys.\  J.,} 745, 51.

\bibitem[Strachan et al.(2002)]{St02}
Strachan, L., Suleiman, R., Panasyuk, A. V., et al. 2002,
``Empirical Densities, Kinetic Temperatures, and Outflow Velocities in
the Equatorial Streamer Belt at Solar Minimum,''
{\em Astrophys.\  J.,} 571, 1008.

\bibitem[Suzuki and Inutsuka(2006)]{SI06}
Suzuki, T. K., and  Inutsuka, S.-I. 2006,
``Solar winds driven by nonlinear low-frequency Alfv\'{e}n waves from
the photosphere: Parametric study for fast/slow winds and
disappearance of solar winds,''
{\em J.\  Geophys.\  Res.,} 111, A06101.

\bibitem[Tanaka et al.(2013)]{Tn13}
Tanaka, Y. A., Suzuki, T. K., and Inutsuka, S.-I. 2013,
``Atmospheric escape by magnetically driven wind from gaseous planets,''
{\em Astrophys.\  J.,} 792, 18.

\bibitem[Threlfall et al.(2013)]{Th13}
Threlfall, J., De Moortel, I., McIntosh, S. W., et al. 2013,
``First Comparison of Wave Observations from CoMP and AIA/SDO,''
{\em Astron.\  Astrophys.,} 556, A124.

\bibitem[Tian et al.(2014)]{Ti14}
Tian, H., DeLuca, E., Cranmer, S. R., et al. 2014,
``Prevalence of Small-scale Jets from the Networks of the
Solar Transition Region and Chromosphere,''
{\em Science,} 346, 1255711.

\bibitem[Tian et al.(2010)]{Ti10}
Tian, H., Tu, C-.Y., Marsch, E., et al. 2010,
``Upflows in the upper transition region of the quiet Sun,''
in {\em Twelfth International Solar Wind Conference,}
AIP Conf.\  Proc.\  1216, (New York: AIP Press), 36.

\bibitem[Title and Schrijver(1998)]{TS98}
Title, A. M., and Schrijver, C. J. 1998,
``The Sun's magnetic carpet,''
in {\em 10th Cambridge Workshop on Cool Stars, Stellar Systems and
the Sun,} ASP Conf.\  Series 154, ed. R. Donahue and J. Bookbinder
(San Francisco: Astron.\  Soc.\  Pacific), 345.

\bibitem[Tomczyk et al.(2013)]{To13}
Tomczyk, S., Gallagher, D., Wu, Z., et al. 2013,
``The COronal Solar Magnetism Observatory (COSMO) Large Aperture
Coronagraph,''
{\em Geophys.\  Res.\  Abstracts,} 15, EGU2013-12573.

\bibitem[van Ballegooijen et al.(2011)]{vB11}
van Ballegooijen, A. A., Asgari-Targhi, M., Cranmer, S. R., and
DeLuca, E. 2011,
``Heating of the Solar Chromosphere and Corona by Alfv\'{e}n Wave
Turbulence,''
{\em Astrophys. J.,} 736, 3.

\bibitem[van Ballegooijen et al.(2014)]{vB14}
van Ballegooijen, A. A., Asgari-Targhi, M., and Berger, M. A. 2014,
``On the Relationship Between Photospheric Footpoint Motions and
Coronal Heating in Solar Active Regions,''
{\em Astrophys. J.,} 787, 87.

\bibitem[van Ballegooijen and Cranmer(2008)]{vB08}
van Ballegooijen, A. A., and Cranmer, S. R. 2008,
``Hyperdiffusion as a Mechanism for Solar Coronal Heating,''
{\em Astrophys. J.,} 682, 644.

\bibitem[van Ballegooijen et al.(1998)]{vB98}
van Ballegooijen, A. A., Nisenson, P., Noyes, R. W., et al. 1998,
``Dynamics of Magnetic Flux Elements in the Solar Photosphere,''
{\em Astrophys. J.,} 509, 435.

\bibitem[V\'{a}squez and Raymond(2005)]{VR05}
V\'{a}squez, A. M., and Raymond, J. C. 2005,
``Oxygen abundance in coronal streamers,''
{\em Astrophys. J.,} 619, 1132.

\bibitem[Velli(1993)]{Ve93}
Velli, M. 1993,
``On the propagation of ideal, linear Alfv\'{e}n waves in radially
stratified stellar atmospheres and winds,''
{\em Astron.\  Astrophys.,} 270, 304.

\bibitem[Velli et al.(1989)]{Ve89}
Velli, M., Grappin, R., and Mangeney, A. 1989,
``Turbulent cascade of incompressible unidirectional Alfv\'{e}n waves
in the interplanetary medium,''
{\em Phys.\  Rev.\  Lett.,} 63, 1807.

\bibitem[Velli et al.(1991)]{Ve91}
Velli, M., Grappin, R., and Mangeney, A. 1991,
``Waves from the Sun?''
{\em Geoph.\  Astrophys.\  Fluid Dyn.,} 62, 101.

\bibitem[Ventura et al.(2005)]{Ve05}
Ventura, R., Spadaro, D., Cimino, G., et al. 2005,
``Streamers and adjacent regions observed by UVCS/SOHO: A comparison
between different phases of solar activity,''
{\em Astron.\  Astrophys.,} 430, 701.

\bibitem[Wang(1994)]{Wa94}
Wang, Y.-M. 1994, ``Polar plumes and the solar wind,''
{\em Astrophys.\  J.,} 435, L153.

\bibitem[Wang et al.(2012)]{Wa12}
Wang, Y.-M., Grappin, R., Robbrecht, E., et al. 2012,
``On the Nature of the Solar Wind from Coronal Pseudostreamers,''
{\em Astrophys.\  J.,} 749, 182.

\bibitem[Wang and Sheeley(1990)]{WS90}
Wang, Y.-M., and Sheeley, N. R., Jr., 1990,
``Solar wind speed and coronal flux-tube expansion,''
{\em Astrophys.\  J.,} 355, 726.

\bibitem[Wiegelmann et al.(2005)]{Wi05}
Wiegelmann, T., Lagg, A., Solanki, S. K., et al. 2005,
``Comparing magnetic field extrapolations with measurements of
magnetic loops,''
{\em Astron.\  Astrophys.,} 433, 701.

\bibitem[Wilhelm et al.(2011)]{Wi11}
Wilhelm, K., Abbo, L., Auch\`{e}re, F., et al. 2011,
``Morphology, dynamics and plasma parameters of plumes and inter-plume
regions in solar coronal holes,''
{\em Astron.\  Astrophys.\  Review,} 19, 35.

\bibitem[Withbroe and Noyes(1977)]{WN77}
Withbroe, G. L., and Noyes, R. W. 1977,
``Mass and energy flow in the solar chromosphere and corona,''
{\em Ann.\  Rev.\  Astron.\  Astrophys.,} 15, 363.

\bibitem[Woolsey and Cranmer(2014)]{WC14}
Woolsey, L. N., and Cranmer, S. R. 2014,
``Turbulence-driven Coronal Heating and Improvements to Empirical
Forecasting of the Solar Wind,''
{\em Astrophys.\  J.,} 787, 160.

\bibitem[Yang et al.(2014)]{Ya14}
Yang, Y.-F., Qu, H.-X., Ji, K.-F., et al. 2014,
``Characterizing motion types of G-band bright points in the quiet Sun,''
{\em Research in Astron.\  Astrophys.,} in press, arXiv:1407.7958.

\end{thebibliography}

\end{document}